\documentclass[orivec]{llncs}

\usepackage[utf8]{inputenc}
\usepackage[T1]{fontenc}

\usepackage{amsmath,amssymb}
\usepackage{graphicx}
\usepackage{xcolor}
\usepackage{array}
\usepackage{etoolbox}
\usepackage{hyperref}
\usepackage{subcaption}
\usepackage{tikz}
\usepackage{fancyvrb}
\usepackage{pifont}
\usepackage{xspace}
\usepackage{rotating}
\usepackage{booktabs}
\usepackage{microtype}

\usepackage{todonotes}
\usepackage[normalem]{ulem}
\usepackage{MnSymbol}
\usepackage[autolanguage]{numprint}

\usepackage{multirow}
\usetikzlibrary{shapes}
\usetikzlibrary{plotmarks}
\usepackage{tabto}
\usepackage{listings}
\usepackage{colortbl}
\usepackage[inline,shortlabels]{enumitem}
\usepackage{circledsteps}
\pgfkeys{/csteps/outer color=black}
\pgfkeys{/csteps/fill color=ldbcpale}
\usepackage[most]{tcolorbox}

\newtoggle{anonymizesystems}

\newcommand{\yedscale}{0.48}

\newcommand{\ldbcsnb}{LDBC SNB\xspace}
\newcommand{\snb}{SNB\xspace}
\newcommand{\snbbi}{SNB BI\xspace}
\newcommand{\interactivevone}{Interactive~v1\xspace}
\newcommand{\interactivevtwo}{Interactive~v2\xspace}
\newcommand{\snbinteractive}{SNB Interactive\xspace}
\newcommand{\snbinteractivevone}{SNB Interactive~v1\xspace}
\newcommand{\snbinteractivevtwo}{SNB Interactive~v2\xspace}

\newcommand{\biw}{BI workload\xspace}

\newcommand{\aka}{a.k.a.\xspace}
\newcommand{\eg}{e.g.\xspace}
\newcommand{\ie}{i.e.\xspace}

\newcommand{\etc}{etc.\xspace}
\newcommand{\vs}{vs.\xspace}

\definecolor{ldbcpale}{HTML}{86cd7d}
\definecolor{ldbc}{HTML}{439539}

\definecolor{mydarkyellow}{HTML}{ffc329}
\definecolor{mylightyellow}{HTML}{fee090}

\definecolor{mydarkblue}{HTML}{80d6ed}
\definecolor{mylightblue}{HTML}{e0f3f8}

\definecolor{red}{rgb}{0.7, 0.11, 0.11}
\definecolor{blue}{rgb}{0.0, 0.0, 0.55}
\definecolor{green}{rgb}{0.0, 0.42, 0.24}
\definecolor{grey}{rgb}{0.52, 0.52, 0.51}

\newcommand{\type}[1]{\textsf{#1}}

\newcommand{\Comment}{\type{Comment}\xspace}
\newcommand{\Comments}{\type{Comments}\xspace}
\newcommand{\ContainerOf}{\type{containerOf}\xspace}
\newcommand{\Forum}{\type{Forum}\xspace}
\newcommand{\Forums}{\type{Forums}\xspace}

\newcommand{\Message}{\type{Message}\xspace}
\newcommand{\Messages}{\type{Messages}\xspace}
\newcommand{\Person}{\type{Person}\xspace}

\newcommand{\Persons}{\type{Persons}\xspace}
\newcommand{\Post}{\type{Post}\xspace}
\newcommand{\Posts}{\type{Posts}\xspace}
\newcommand{\Tag}{\type{Tag}\xspace}

\newcommand{\University}{\type{University}\xspace}

\newcommand{\Country}{\type{Country}\xspace}
\newcommand{\Countries}{\type{Countries}\xspace}

\newcommand{\rootPost}{\type{rootPost}\xspace}

\newcommand{\hasCreator}{\type{hasCreator}\xspace}
\newcommand{\hasMember}{\type{hasMember}\xspace}

\newcommand{\knows}{\type{knows}\xspace}

\newcommand{\snbperson}{\texttt{person}\xspace}
\newcommand{\snbfriend}{\texttt{friend}\xspace}
\newcommand{\snbcomment}{\texttt{comment}\xspace}

\newcommand{\OneK}{\numprint{1000}\xspace}
\newcommand{\ThreeK}{\numprint{3000}\xspace}
\newcommand{\TenK}{\numprint{10000}\xspace}
\newcommand{\ThirtyK}{\numprint{30000}\xspace}

\newcommand{\tpcH}{\mbox{TPC-H}\xspace}

\definecolor{Person}{HTML}{fdb462}
\definecolor{Message}{HTML}{bebada}
\definecolor{Forum}{HTML}{b3de69}
\definecolor{Comment}{HTML}{80b1d3}
\definecolor{Post}{HTML}{fb8072}
\definecolor{Company}{HTML}{ccebc5}
\definecolor{University}{HTML}{ffed6f}
\definecolor{City}{HTML}{8dd3c7}
\definecolor{Tag}{HTML}{fccde5}
\definecolor{Country}{HTML}{ffffb3}

\definecolor{grey}{rgb}{0.52, 0.52, 0.51}
\definecolor{red}{rgb}{0.7, 0.11, 0.11}
\definecolor{blue}{rgb}{0.0, 0.0, 0.55}
\definecolor{green}{rgb}{0.0, 0.42, 0.24}

\usetikzlibrary{patterns}

\newcolumntype{Y}{>{\raggedright\arraybackslash}X}
\newcolumntype{R}[1]{>{\raggedleft\let\newline\\\arraybackslash\hspace{0pt}}m{#1}}
\newcolumntype{C}[1]{>{\centering\let\newline\\\arraybackslash\hspace{0pt}}m{#1}}

\usetikzlibrary{patterns}
\def\shadedBox(#1,#2,#3){
  \fill[pattern=north west lines,pattern color=grey] (#1,#2) --  (#1,#2 - #3) -- (#1 + 0.3,#2 - #3) --  (#1 + 0.3,#2);
   \draw [grey,thin,dashed] (#1,#2)  -- (#1,#2 - #3);
   \draw [grey,thin,dashed] (#1 + 0.3,#2) -- (#1 + 0.3,#2 - #3);
   \draw [grey,line width=0.6mm] (#1,#2 - #3) -- node[midway,below,grey] {$\Delta$} (#1 + 0.3,#2 - #3);`'
}

\let\oldDelta\Delta
\renewcommand{\Delta}{\mathrm{\oldDelta}}

\tcbset{My Title Style Special/.style={
    arc=0pt,outer arc=0pt, %
    boxsep=0pt,
    boxrule=0.5pt,
    top=2pt,       
    bottom=2pt,
    left=5pt, %
    right=5pt, %
    leftrule=4pt,
    rightrule=1pt,
    colframe=black,
    standard jigsaw,
    opacityback=0,
}}

\newenvironment{benchmarkDesignCriterion}{\begin{tcolorbox}[My Title Style Special]}{\end{tcolorbox}}

\newcommand{\snbOperationInTitle}[1]{{\ttfamily \fontseries{b}\selectfont #1}\xspace}
\newcommand{\INSInTitle}[1][]{\snbOperationInTitle{INS#1}}
\newcommand{\DELInTitle}[1][]{\snbOperationInTitle{DEL#1}}
\newcommand{\CRInTitle}[1][]{\snbOperationInTitle{CR#1}}
\newcommand{\SRInTitle}[1][]{\snbOperationInTitle{SR#1}}

\newcommand{\snbOperation}[1]{{\ttfamily \fontseries{m}\selectfont #1}\xspace}
\newcommand{\INS}[1][]{\snbOperation{INS#1}}
\newcommand{\DEL}[1][]{\snbOperation{DEL#1}}
\newcommand{\CR}[1][]{\snbOperation{CR#1}}
\newcommand{\SR}[1][]{\snbOperation{SR#1}}

\renewcommand{\paragraph}[1]{\noindent\uline{\textbf{#1.}}\xspace}

\usepackage{listings}

\usepackage{textcomp}
\definecolor{keyword}{HTML}{2771a3}
\definecolor{pattern}{HTML}{b53c2f}
\definecolor{string}{HTML}{be681c}
\definecolor{relation}{HTML}{7e4894}
\definecolor{variable}{HTML}{107762}
\definecolor{comment}{HTML}{8d9094}

\lstdefinelanguage{cypher}
{
	morekeywords={
		MATCH, OPTIONAL, WHERE, NOT, AND, OR, XOR, RETURN, DISTINCT, ORDER, BY, ASC, ASCENDING, DESC, DESCENDING, UNWIND, AS, UNION, WITH, ALL, CREATE, DELETE, DETACH, REMOVE, SET, MERGE, SET, SKIP, LIMIT, IN, CALL, CASE, WHEN,
		INDEX, DROP, UNIQUE, CONSTRAINT, EXPLAIN, PROFILE, START, FOREACH, %
		GROUP, HAVING,
	},
	sensitive=true,
	morecomment=[l]{//},
	morecomment=[s]{/*}{*/},
	morestring=[b]{"},
	literate=*{<<}{\guillemotleft{}}{1}{>>}{\guillemotright{}}{1},
}

\newcommand{\mycdots}{\cdot\!\cdot\!\cdot}
\begin{document}

\title{The LDBC Social Network Benchmark Interactive Workload v2: A Transactional Graph Query Benchmark with Deep Delete Operations}

\author{%
    David P\"{u}roja\inst{1} \and
    Jack Waudby\inst{2} \and
    Peter Boncz\inst{1} \and
    Gábor Szárnyas\inst{1}
}
\institute{
    \textsuperscript{1}~CWI, the Netherlands,
    \textsuperscript{2}~Newcastle University, School of Computing\\
    \email{david.puroja@ldbcouncil.org},
    \email{j.waudby2@newcastle.ac.uk},
    \email{boncz@cwi.nl},
    \email{gabor.szarnyas@ldbcouncil.org}
}

\maketitle

\begin{abstract}
The LDBC Social Network Benchmark's Interactive workload captures an OLTP scenario operating on a correlated social network graph. It consists of complex graph queries executed concurrently with a stream of updates operation. Since its initial release in 2015, the Interactive workload has become the de facto industry standard for benchmarking transactional graph data management systems. As graph systems have matured and the community's understanding of graph processing features has evolved, we initiated the renewal of this benchmark. This paper describes the draft Interactive v2 workload with several new features:
delete operations, a cheapest path-finding query, support for larger data sets, and a novel temporal parameter curation algorithm that ensures stable runtimes for path queries.

\end{abstract}

\section{Introduction}
\label{sec:introduction}

\paragraph{LDBC}
The Linked Data Benchmark Council (LDBC)%
\footnote{\url{https://ldbcouncil.org}}
is a non-profit organization dedicated to designing benchmarks for graph data management~\cite{DBLP:journals/vldb/SahuMSLO20,DBLP:journals/cacm/SakrBVIAAAABBDV21}.
LDBC has strong industrial participation in the form of 21~companies, including database, hardware, and cloud vendors.
Its membership also includes 3~non-commercial institutions and 60\texttt{+} individual members.
LDBC acts as an independent authority for benchmarks and oversees the use of its benchmarks through a stringent auditing process.
Thanks to this, audited LDBC benchmark results allow quantitative and objective comparison of different technological solutions, which is expected to stimulate progress through competition.
Next, we describe the two main workloads of the LDBC SNB suite.

\paragraph{SNB Interactive v1 workload}
The LDBC Social Network Benchmark (SNB) \interactivevone workload was published in 2015~\cite{DBLP:conf/sigmod/ErlingALCGPPB15}.
It is a transactional benchmark that targets OLTP data management systems with graph features (\eg path-finding).
\snbinteractive has been influential in the graph data management space:
as of August 2023, 24~audited results were published using this workload.%
\footnote{\url{https://ldbcouncil.org/benchmarks/snb-interactive/}}

\paragraph{SNB Business Intelligence workload}
The LDBC SNB Business Intelligence (BI) workload was released in 2022~\cite{DBLP:journals/pvldb/SzarnyasWSSBWZB22}.
This workload uses an improved data generator, which introduces support for delete operations~\cite{DBLP:conf/sigmod/WaudbySPS20} and scale factors up to SF\ThirtyK.
The workload captures an OLAP scenario with heavy-hitting analytical queries that touch on large portions of the graph (\eg \Messages created within a 100-day period or \Persons living in China) and applies daily batches of updates.
It targets both DBMSs and data analytical systems such as Spark.

\paragraph{Motivation}
As of 2023, more than 8~years passed since the \snbinteractivevone workload's release.
Therefore, we decided to renew it to ensure its continued relevance.
The new version's key novel features are improved scalability, coverage of cheapest path-finding, and inclusion of delete operations.
The first two new features were part of the natural evolution of the benchmark.
The decision to support delete operations was motivated by a number of factors.
On the technical side, running the workload's complex queries efficiently while applying delete operations assumes mature transaction support that is now expected by users of graph(-capable) DBMSs.
Deletes also make certain graph algorithms, such as cheapest paths, more difficult to compute incrementally~\cite{DBLP:conf/esa/RodittyZ04},
thus limiting the effects of caching and incremental view maintenance.
On the business side, supporting deletes is necessitated by law in several jurisdictions,
exemplified by the EU's General Data Protection Regulation (GDPR)~\cite{DBLP:journals/pvldb/ShastriBWKC20}.

\begin{figure}[htb]
    \centering
    \vspace{-3.5ex}
    \includegraphics[scale=\yedscale]{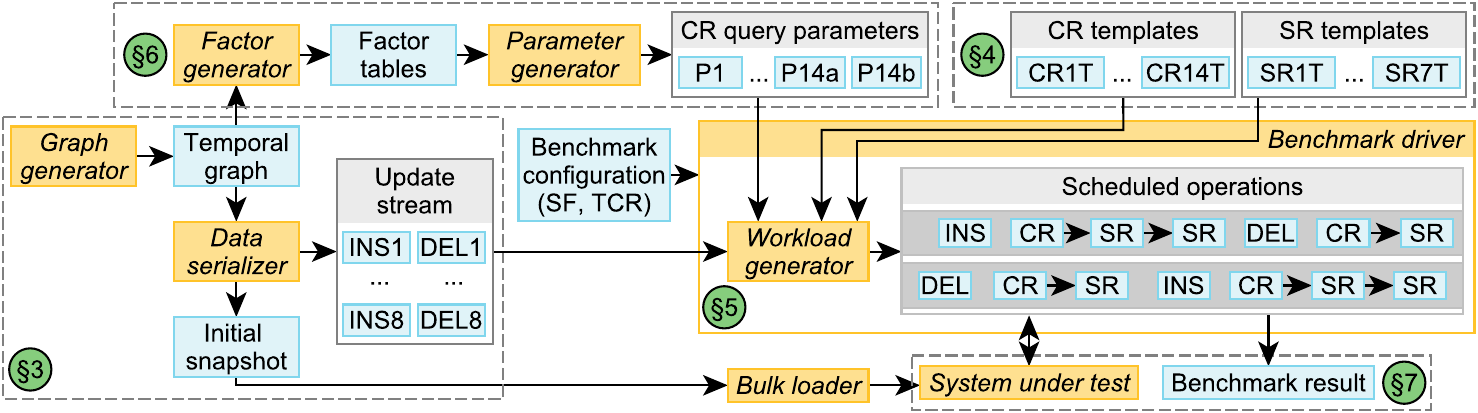}
    \caption{
        Components and workflow of the Interactive~v2 workload.
        The corresponding sections are shown in green circles \Circled{\textsf{\scriptsize \S}}.
        Legend:
        \fcolorbox{mydarkyellow}{mylightyellow}{{\scriptsize \textit{\textsf{Software component}}}}
        \fcolorbox{mydarkblue}{mylightblue}{{\scriptsize \textsf{Data artifact\vphantom{p}}}}
    }
    \label{fig:interactive-components}
    \vspace{-3.5ex}
\end{figure}

\paragraph{Contributions and paper structure}
This paper presents the updated \snbinteractivevtwo workload following the workflow shown in \autoref{fig:interactive-components}.
Interestingly, while all three new features (scalability, cheapest path-finding, and delete operations)
were already supported in the \biw, adopting them into the highly transactional, concurrent \snbinteractivevtwo workload presented several complex technical challenges.
We document the challenges and our key design principles in \autoref{sec:design-principles}.
We present the SNB data set in \autoref{sec:data}, the benchmark's operations in \autoref{sec:operations}, and the workload in \autoref{sec:workload-and-driver}.
In \autoref{sec:parameter-curation}, we introduce the parameter generator's novel extensions that were necessitated by delete operations.
In \autoref{sec:framework}, we discuss how the benchmark is used in practice.
We discuss LDBC's other transactional benchmark, the FinBench, in \autoref{sec:related-work}.
\autoref{sec:conclusion} summarizes our contributions and outlines future directions.

\section{Design principles}
\label{sec:design-principles}

During LDBC's benchmark design process, we follow the Benchmark Handbook~\cite{Gray:BenchmarkBook}, which prescribes four criteria for domain-specific benchmarks:
\begin{enumerate*}[(1)]
    \item \emph{relevance},
    \item \emph{portability},
    \item \emph{scalability},
    \item \emph{simplicity}.
\end{enumerate*}
In the following, we discuss the \snbinteractivevtwo workload's approach for complying with these criteria.

\subsection{Relevance: Choke point-based design process and domain}
\label{sec:relevance}
\label{sec:choke-points}

\begin{benchmarkDesignCriterion}
The benchmark must measure the peak performance of systems when performing typical operations within the target problem domain.
\end{benchmarkDesignCriterion}

\paragraph{Choke points}
To ensure \emph{relevance}, LDBC's benchmark design process uses \emph{choke points}~\cite{DBLP:conf/tpctc/BonczNE13},
\ie technical difficulties that are known to be challenging for the present generation of DBMSs.
Choke points are identified by expert data systems architects and are also influenced by the input from users of graph data management systems who contribute their use cases at LDBC's Technical User Community meetings%
\footnote{\url{https://ldbcouncil.org/tags/tuc-meeting/}}.
The initial choke points of \snbinteractive were based on the influential \tpcH benchmark~\cite{tpch} benchmark
and were later extended with choke points that target graph-specific features such as cardinality estimation for paths and the execution of path-finding queries.
LDBC workloads are designed using an iterative process to ensure full coverage of the choke points required for a given workload category.

\paragraph{Social network domain}
The \ldbcsnb uses the \emph{social network} domain because its concepts (\Person, \Forum, \Message, \etc) are well-understood.
Moreover, the social network domain makes it easy to reason about some of the interesting phenomena captured in the choke points.
For example, the power law distribution and correlations (\autoref{sec:correlations}) observable in real-life (social) networks trigger the challenges for cardinality estimation.

\subsection{Portability: Implementation rules}
\label{sec:portability}

\begin{benchmarkDesignCriterion}
The benchmark should be implementable on different systems/architectures.
\end{benchmarkDesignCriterion}

\noindent
The \snbinteractivevtwo workload guarantees \emph{portability} by taking an agnostic stance on implementation details.

\paragraph{Data model}
Implementations are allowed to use any data model, including the property graph, RDF, and relational models.
They are also free to choose their input format for bulk loading (\eg CSV, \mbox{N-Triples}).

\paragraph{Implementation language}
Implementations may use
declarative query languages (SQL, Cypher~\cite{DBLP:conf/sigmod/FrancisGGLLMPRS18}, GQL, SQL/PGQ~\cite{DBLP:conf/sigmod/DeutschFGHLLLMM22}, \etc)
or
general-purpose imperative programming languages (C\texttt{++}, Java, \etc).
However, results in these two categories are ranked on separate leaderboards as the latter systems have a significant advantage due to their use of hand-coded highly-optimized query plans.

\paragraph{Setup}
There are no restrictions on the operating system, hardware architecture, or number of machines used
(both single-node and distributed setups are allowed).

\subsection{Scalability: Scalable data generator and driver}
\label{sec:scalability}

\begin{benchmarkDesignCriterion}
The benchmark should apply to small and large computer systems.
\end{benchmarkDesignCriterion}

\noindent
Improving \emph{scalability} was a key goal during the design of \snbinteractivevtwo.
While we could leverage the improved data generator of the BI workload~\cite{DBLP:journals/pvldb/SzarnyasWSSBWZB22}, scaling the \emph{workload execution} posed additional challenges.
By its nature, simulating a transactional database workload requires highly concurrent execution of the operations.%
\footnote{Most audited \interactivevone implementations use 48~read and 32/64~write threads.}
This requires the operations in the workload to be partitioned, which is a major challenge as most of the \Persons in the social network belong to a single connected component that does not lend itself to any na\"ive partitioning strategy.
Moreover, update operations often have long dependency chains that need to be tracked,
\eg a friendship can only be deleted if it already exists, the creation of a friendship requires both \Persons to exist, \etc
Therefore, simulating a transactional graph processing scenario is not possible using on-the-fly workload generation techniques commonly employed in database benchmarks.
Instead, \snbinteractivevtwo requires extensive offline data, update stream (\autoref{sec:graph-generation-stages}), and parameter generation steps (\autoref{sec:parameter-curation}) prior to the benchmark.

\subsection{Simplicity: Stable query runtimes, single output metric}
\label{sec:simplicity}

\begin{benchmarkDesignCriterion}
The benchmark must be understandable and its results must be easy to interpret, otherwise, it lacks credibility.
\end{benchmarkDesignCriterion}

\paragraph{Stable runtimes}
To make the benchmark results easy to interpret,
it is desirable that instances of a given query type have similar expected runtimes (referred to as \emph{stable runtimes}).
Ensuring this for graph workloads is non-trivial due to the highly skewed distribution exhibited in real-world networks~\cite{newman_power_law}.
For example, in a social network, a few \Person nodes have a very large number of edges while others only have a few connections.
This has a significant impact on runtimes: if query parameters are selected using uniform random sampling, query runtimes will be unstable, often exhibiting a multimodal distribution that spreads across many orders of magnitude and has several outliers~\cite{DBLP:conf/tpctc/GubichevB14,DBLP:journals/pvldb/SzarnyasWSSBWZB22}.
\snbinteractivevtwo employs a sophisticated \emph{parameter curation} process to select input parameters that ensure stable runtimes (\autoref{sec:parameter-curation}).

\paragraph{Guaranteed executability}
Stable runtimes also necessitate that operations are \emph{executable} at their scheduled start time.
For example, if an operation targets entities that do not yet exist or were already deleted, the operation becomes trivial or results in a runtime exception, compromising stable runtimes.
Therefore, our workload generator ensures that its operations are always executable.

\paragraph{Single metric}
The result of an \snbinteractive benchmark run is characterized by a single metric, \emph{throughput} (operations/second), which captures the system-under-test's end-to-end performance on a transactional graph workload.

\section{Data sets}
\label{sec:data}

The LDBC SNB workloads include a scalable distributed data generator based on Spark.%
\footnote{\url{https://github.com/ldbc/ldbc_snb_datagen_spark}}
Here, we give an overview of the data sets used in the benchmark.

\subsection{Graph schema}
\label{sec:graph-schema}

\begin{figure}[htb]
    \vspace{-4.5ex}
    \centering
    \includegraphics[scale=\yedscale]{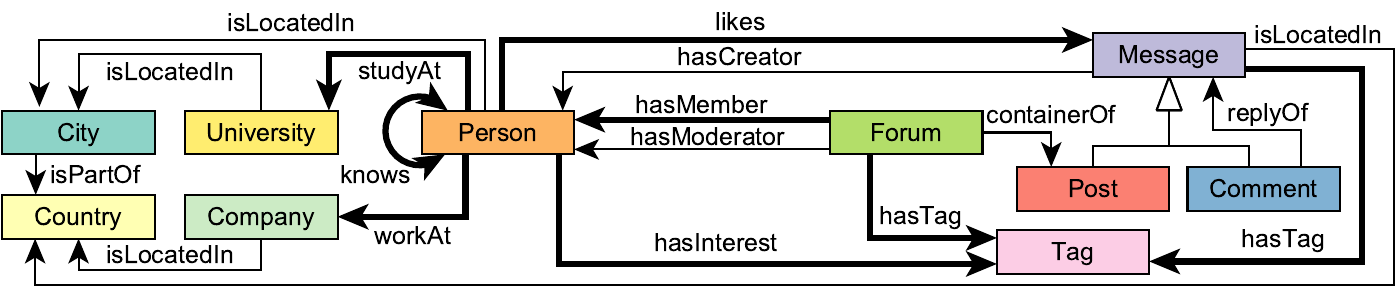}
    \caption{
        A subset of the LDBC SNB graph schema visualized using a UML-like notation.
        Thick lines denote many-to-many relationships.
    }
    \label{fig:schema}
    \vspace{-3.5ex}
\end{figure}

\noindent
The graph schema of \ldbcsnb has 14~node types connected by 20~edge types.
The data set consists of a \Person--\knows--\Person (friendship) graph and a number of \Message threads within \Forums.
The root of a \Message thread is a \Post and the rest of the thread consists of \Comments.
All \Messages are connected to \Persons by creatorship and likes edges.
A simplified schema is shown in \autoref{fig:schema}.

\subsection{Distribution and correlations}
\label{sec:correlations}

The data set contains two types of graph-shaped data structures.
First, the \Message threads form trees and constitute the majority of the data.
Second, the \Person--\knows--\Person subgraph is a network with many-to-many relationships whose distribution is modelled after Facebook~\cite{facebook_anatomy}
with the social graph exhibiting the small-world phenomenon~\cite{watts_strogatz_98} characterized by a small diameter.

A unique feature not observed in other data generators is that the attribute distributions are skewed and correlate both within an entity (\eg people living in France have predominantly French names).
Moreover, the graph has structural correlations: following the \emph{homophily principle}~\cite{mcpherson2001birds}, people are more likely to be friends if they studied at the same \University at the same time, live in close proximity, and/or have the same interests.
These correlations are exploited by the workload to stress choke points for querying correlated data (see \autoref{sec:cr3-variants}).

\subsection{Graph generation stages}
\label{sec:graph-generation-stages}
\label{def:cutoff-date}

\paragraph{Temporal graph}
The data generator first produces a \emph{temporal graph}, which contains all entities that exist at some point in the simulated social network's 3-year time period,
\ie between Jan 1, 2010 and Dec 31, 2012.
During this time, entities are inserted and deleted in the network, and the timestamps of these events are captured according to their time of occurrence in the \emph{simulation time}.
The insertion and deletion of entities follow realistic time intervals and conform to the semantics of the social network.
Namely:
\begin{enumerate*}[(1)]
    \item
        The deletion dates of \Persons are based on the statistics collected from the collapse of a real-world social network~\cite{DBLP:journals/socnet/LorinczKGT19}.
        When a \Person is deleted, the content they created is also deleted~\cite{DBLP:conf/tpctc/WaudbySKMBS20}.
    \item
        The network contains infrequent flashmob events such as spikes in insertions of \Messages for a given \Tag~\cite{DBLP:conf/sigmod/ErlingALCGPPB15}.
\end{enumerate*}
Deep delete operations and flashmob events are unique to the \ldbcsnb data generator:
according to a recent survey~\cite{DBLP:journals/csur/BonifatiHPS20}, these features are not supported by any other (graph) data generator.

\paragraph{Initial snapshot and update stream}
As the second step in the data generation, the \emph{data serializer} splits the temporal graph into two parts by setting a \emph{cutoff date} at 97\% of the simulation time (Nov 29, 2012).
The entities created before the cutoff date form the \emph{initial snapshot},
while the entity creations and deletions occurring %
after the cutoff date form the \emph{update stream}.

\subsection{Scale factors}
\label{sec:scale-factors}

\begin{table}[htb]
    \vspace{-2.5ex}
    \centering
    \setlength{\tabcolsep}{3.2pt}
    \caption{
        SNB Interactive v2 data sets.
        \emph{k:} thousand,
        \emph{M:} million,
        \emph{B:} billion.
    }
    \label{tab:data-set-sizes}
    \begin{tabular}{@{}lrrrrrrrr@{}}
        \toprule
        \bf Scale Factor (SF)
                                        &
        \multicolumn{1}{c}{\bf 10}      &
        \multicolumn{1}{c}{\bf 30}      &
        \multicolumn{1}{c}{\bf 100}     &
        \multicolumn{1}{c}{\bf 300}     &
        \multicolumn{1}{c}{\bf \OneK}   &
        \multicolumn{1}{c}{\bf \ThreeK} &
        \multicolumn{1}{c}{\bf \TenK}   &
        \multicolumn{1}{c}{\bf \ThirtyK}
        \\ \midrule
        \#nodes                         & 27M   & 78M  & 255M & 738M  & 2.4B  & 7.2B  & 23B  & 82.76B  \\
        \#edges                         & 170M  & 506M & 1.7B & 5.1B  & 17B   & 51.9B & 173B & 340.5B \\
        \midrule
        \#\Person nodes                 & 68k   & 170k & 473k & 1.2M  & 3.4M  & 9M    & 26M  & 77M  \\
        \#\knows edges                  & 1.8M  & 5.5M & 19M  & 55.7M & 187M  & 559M  & 1.9B & 6.8B   \\
        \midrule
        \#insert operations             & 44.6M & 127M & 399M & 1.1B  & 3.3B  & 8.9B  & 27B  & 76.7B  \\
        \#delete operations             & 353k  & 1M   & 3.3M & 9.3M  & 28.9M & 79.7M & 245M & 721.8M \\
        \bottomrule
    \end{tabular}
    \vspace{-3.5ex}
\end{table}

\noindent
The data generator produces dynamic social network graphs in different sizes, characterized by \emph{scale factors} (SF) which correspond to the data set's disk usage when serialized in CSV (comma-separated values) format, measured in GiB.
The data generator used for \interactivevone only supports data sets up to SF\OneK.
To improve scalability, \interactivevtwo uses the new Spark-based generator, which was optimized extensively,%
\footnote{For details on the optimization steps, see \url{https://ldbcouncil.org/tags/datagen/}.}
allowing it to scale up to SF\ThirtyK.
\autoref{tab:data-set-sizes} shows the main statistics of the data sets.

\section{Operations}
\label{sec:operations}

The LDBC \snbinteractivevtwo workload uses four types of operations.
There are 14~complex (\CR) and 7~short read queries (\SR).
Update operations include
8~inserts (\INS) and, newly introduced in the \interactivevtwo workload, 8~deletes (\DEL).
The workload mix consists of approximately
8\% \CR,
72\% \SR,
20\% \INS, and
0.2\% \DEL
operations.
In this section, we describe the four operation types using examples.
We also give ranges on how long operations are expected to take in state-of-the-art systems (\autoref{sec:auditing}).

\subsection{Complex read queries (\CRInTitle)}
\label{sec:complex-read-queries}

Complex read queries \CR[1]--\CR[12] discover a given \Person's social environment (one- to three-hop neighbourhoods)
and retrieve related content (\Forums, \Messages, \etc). Queries \CR[13] and \CR[14] perform path-finding between pairs of \Persons.
The runtimes of complex read queries are typically between 1 and 500~ms, making them feasible to compute interactively, in line with the workload's name.

\begin{figure}[htb]
    \vspace{-2.5ex}
    \centering
    \begin{subfigure}{0.61\linewidth}
        \centering
        \includegraphics[scale=\yedscale]{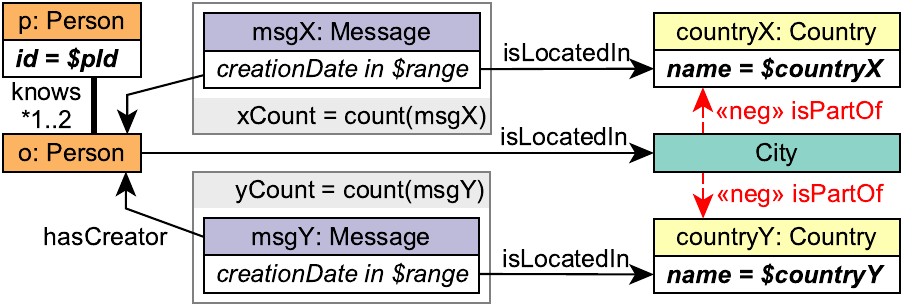}
        \caption{Complex read query \CR[3].}
        \label{fig:interactive-complex-read-03}
    \end{subfigure}
    \begin{subfigure}{0.13\linewidth}
        \centering
        \advance\leftskip0.25cm
        \includegraphics[scale=\yedscale]{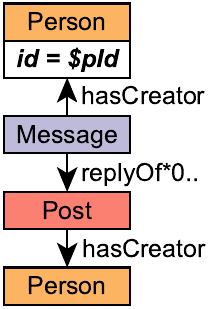}
        \caption{\SR[2].}
        \label{fig:interactive-short-read-02}
    \end{subfigure}
    \begin{subfigure}{0.14\linewidth}
        \centering
        \advance\leftskip0.35cm
        \includegraphics[scale=\yedscale]{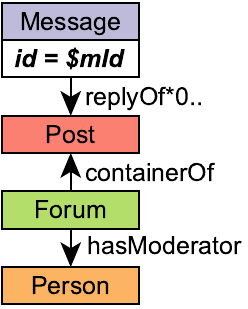}
        \caption{\SR[6].}
        \label{fig:interactive-short-read-06}
    \end{subfigure}
    \caption{Graph patterns of complex and short read queries.}
    \label{fig:read-queries}
    \vspace{-2.5ex}
\end{figure}

\paragraph{\CRInTitle[3]}
\label{sec:cr3}
For a given \Person, find their \emph{friends} and \emph{friends of friends}, who created \Messages in both \Country \texttt{\$countryX} and \texttt{\$countryY} within a given time period.
Only consider \Persons that are foreign to both of those \Countries.
Return the number of their \Messages per \Country, \texttt{xCount} and \texttt{yCount} %
(\autoref{fig:interactive-complex-read-03}).

\paragraph{\CRInTitle[13]}
Return the length of the (unweighted) shortest path between two \Persons.

\paragraph{Cheapest path-finding}
While we strived to keep the changes to the queries minimal, we replaced \CR[14] due to two reasons.
First, we found the original query in \interactivevone to be ill-suited to the workload as it required the enumeration of \emph{all shortest paths} between two \Persons, which can be prohibitively expensive on large scale factors.
Second, we introduced a new choke point,
\textsf{CP-7.6}
\emph{Cheapest path-finding,}%
\footnote{
    The term \emph{shortest paths} refers to the problem of finding \emph{unweighted shortest paths}, which can be solved with the BFS algorithm.
    We use \emph{cheapest paths} to refer to the \emph{weighted shortest paths} problem which can be solved using \eg Dijkstra's algorithm.
}
a key computational kernel and a language opportunity for GQL~\cite{DBLP:conf/sigmod/DeutschFGHLLLMM22}.
Therefore, we changed \CR[14] to use \emph{cheapest paths} instead of \emph{all shortest paths}.

\paragraph{\CRInTitle[14 (new)]}
Given two \Persons, find \emph{any cheapest path} in the interaction subgraph.
This graph contains edges from the \Person--\knows--\Person graph where the endpoint \Persons have exchanged at least one \Message (\ie one \Person created a direct \Comment to a \Message of the other \Person).
The weights of \knows edges are integers defined as
$\max(\mathrm{round}( 40 - \sqrt{\textit{numInteractions}} ), 1)$.

\subsection{Short read queries (\SRInTitle)}
\label{sec:short-read-queries}

Short read queries perform local neighbourhood lookups on \Persons and \Messages.
Most short read queries can be evaluated in 0.1 to 75~ms.

\paragraph{\SRInTitle[2]}
Given a start \Person, retrieve their last 10~\Messages.
For each \Message, return it with the root \Post in its thread, and the author of that \Post (\autoref{fig:interactive-short-read-02}).

\paragraph{\SRInTitle[6]}
Given a \Message, retrieve its container \Forum (directly for \Posts, via the root \Post for \Comments) and the \Person that moderates that \Forum (\autoref{fig:interactive-short-read-06}).

\subsection{Insert operations (\INSInTitle)}
\label{sec:insert-operations}

Insert operations add new entities from the update stream to the graph.
A typical insert operation takes between 0.1 and 100~ms.

\paragraph{\INSInTitle[5]}
Insert a \hasMember edge between a \Person and a \Forum.
The executability of this operation depends on the existence of its two endpoint nodes.

\paragraph{\INSInTitle[6]}
Insert a \Post node.
This operation's executability depends on two nodes:
both the \Person creating the \Post and the \Forum containing it must exist.
When the \Post is inserted, they are linked to it via \hasCreator and \ContainerOf edges.

\subsection{Delete operations (\DELInTitle)}
\label{sec:delete-operations}

\begin{figure}[htb]
    \vspace{-2.5ex}
    \centering
    \begin{subfigure}{0.47\linewidth}
        \centering
        \includegraphics[scale=\yedscale]{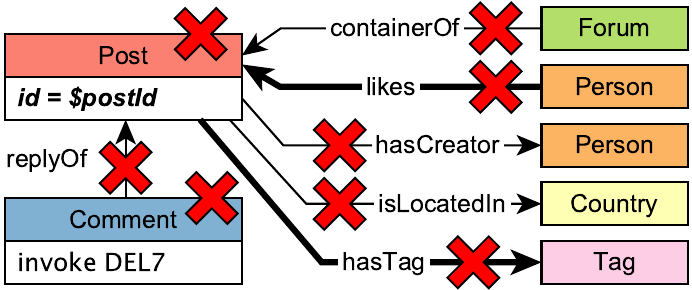}
        \caption{Delete operation \DEL[6]: Remove \Post.}
        \label{fig:delete-06}
    \end{subfigure}
    \begin{subfigure}{0.52\linewidth}
        \centering
        \includegraphics[scale=\yedscale]{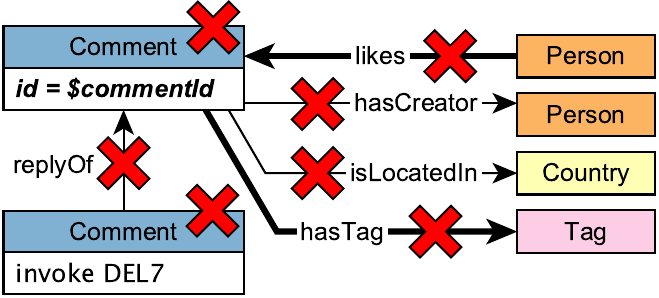}
        \caption{Delete operation \DEL[7]: Remove \Comment.}
        \label{fig:delete-07}
    \end{subfigure}
    \caption{
        Cascading delete operations in Interactive v2.
        Symbol \includegraphics[width=0.25cm]{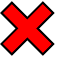} denotes deletion.
    }
    \label{fig:deletes}
    \vspace{-2.5ex}
\end{figure}

The Interactive~v2 workload uses \emph{deep cascading delete operations}.
Cascading deletes capture the behaviour of real social networks where users expect their content to be removed once they delete their accounts.
The technical reasons for requiring cascading delete operations are two-fold:
\begin{enumerate*}[(1)]
    \item
        \textbf{Preventing dangling edges.}
        To maintain the integrity of the graph, it is required there are no dangling edges thus nodes must be always deleted with all their edges.
        To prevent dangling edges, most graph DBMSs support the automatic deletion of edges attached to a given node, \eg through Cypher's \texttt{DETACH DELETE} clause~\cite{DBLP:journals/pvldb/GreenGLLMPSSV19}.
        To achieve the same effect, RDBMSs can make use of \texttt{FOREIGN KEY} constraints with the \texttt{ON DELETE CASCADE} clause.
    \item
        \textbf{Testing triggered deletions.}
        Node deletions can trigger the deletion of other nodes (\autoref{fig:deletes}).
        For example, according to the SNB schema's constraints,
        the deletion of a \Post implies the deletion of all its descendant \Comments along with their edges.
        Such deletions may be implemented using
        triggers,
        constraints (RDBMSs may again harness \texttt{FOREIGN KEY}s),
        or by formulating (potentially recursive) subqueries that determine which other nodes need to be deleted with \texttt{DELETE ... USING} clause.
\end{enumerate*}

\paragraph{Choke points}
The coverage of delete features is ensured by three new choke points:
\textsf{CP-9.3}~\emph{Delete node} (stressed by 4~operations),
\textsf{CP-9.4}~\emph{Delete edge} (8~operations),
and
\textsf{CP-9.5}~\emph{Delete recursively} (4~operations).
In the following, we present two delete operations, both of which cover all new choke points.

\paragraph{\DELInTitle[6]}
Remove a \Post with all its edges and child \Comments via \DEL[7] (\autoref{fig:delete-06}).

\paragraph{\DELInTitle[7]}
Remove a \Comment with all its edges and child \Comments, which are deleted recursively by invoking \DEL[7]
(\autoref{fig:delete-07}).

\section{Workload scheduling and benchmark driver}
\label{sec:workload-and-driver}

In this section, we explain how operations are scheduled in the SNB Interactive workload, how the driver operates, and how the final \emph{throughput} metric is determined.
In all cases, we assume that the system-under-test has been populated with the \emph{initial snapshot} using a \emph{bulk loader} before the driver runs the operations.

\subsection{Scheduling operations}
\label{sec:scheduling}

\paragraph{TCR (total compression ratio)}
The scheduling follows the \emph{simulation time} of the temporal social network graph.
The user-provided \emph{total compression ratio} (TCR) value controls the speed at which the simulation is replayed.
For example, a TCR value of $0.02$ means that the simulation is replayed $50\times$ faster, \ie for every 20~milliseconds in wall clock time, 1~second passes in the simulation time.

\paragraph{Update operations}
The driver replays the update operations starting from the cutoff date (\autoref{def:cutoff-date}), Nov 29, 2012.
The operations are scheduled according to the distance of their start time from this date, adjusted by the TCR.
They are then used to set the cadence of the schedule for the complex reads and, in turn, the short read queries, as we will explain momentarily.

\paragraph{Complex read queries}
The \emph{complex read queries} differ significantly in their expected runtimes as they touch on different amounts of data.
As each query instance contributes equally to the output metric,%
\footnote{Unlike in \tpcH~\cite{tpch} and \snbbi~\cite{DBLP:journals/pvldb/SzarnyasWSSBWZB22}, which use \emph{geometric mean} in their metrics.}
we balance them such that each query type is expected to take the same amount of time to execute.
For example, \CR[14 (new)] is expected to be more difficult than \CR[13], therefore it is scheduled less frequently.
Frequencies vary based on the SF as the relative difficulties of queries change with the data size
(\eg three-hop neighbourhood queries grow faster on larger SFs than one-hop ones).

\paragraph{Short read queries}
Short read queries are triggered by complex read queries and other short read queries, and use their output as their input.
For example, both \CR[3] and \CR[14] trigger \SR[2], which also triggers itself.
This mimics the real-life scenario of a user retrieving more information about \Person profiles based on the result of the earlier queries.
The mapping between complex and short read queries is given in the specification~\cite[Chapter~5]{DBLP:journals/corr/abs-2001-02299}.

\subsection{Driver}
\label{sec:driver}

\paragraph{Driver modes}
The SNB driver has two key modes of operation.
In \emph{cross-validation mode}, %
it tests an implementation against the output of another implementation.
To ensure deterministic results, operations in this mode are executed sequentially with no overlap between queries and updates.
In \emph{benchmark mode} %
the driver performs a benchmark run where queries and updates are issued concurrently from multiple threads.
The run starts with a 30-minute warm-up period, followed by a 2-hour \emph{measurement window}.
This mode does not perform validation as query results may differ (slightly) due to concurrent updates.

\paragraph{Dependency tracking}
To ensure that updates are executable, concurrent threads must be synchronized so that an operation is only executed when its dependencies exist in the network (\eg two \Persons can only become friends if both of them already exist).
This is achieved via maintaining a global clock in the driver and performing \emph{dependency tracking} for the updates~\cite{DBLP:conf/sigmod/ErlingALCGPPB15}: each update operation has a timestamp denoting the creation time of the last operation it depends on.
The data generator calculates these timestamp during generation and ensures that there is a minimum time separation,
$T_\textit{safe}$,
between dependent entities to reduce synchronization overhead in the driver when executing operations.
The driver then only needs to check every $T_\textit{safe}$ time whether a given update operation can be executed.
By default, $T_\textit{safe}$ is set to 10 seconds in the simulation time.

\paragraph{Latency requirements}
The workload simulates a highly transactional scenario where operations are subject to (soft) latency requirements.
To incorporate this in the workload, it prescribes the \emph{95\% on-time requirement}:
for a benchmark run to be successful, 95\% of the operations must start \emph{on-time}, \ie within 1~second
of their scheduled start time.
Benchmark runs where the system-under-test falls behind too much from the schedule are considered invalid.

\paragraph{Throughput}
The throughput of a run is the total number of operations
(\CR, \SR, \INS, \DEL)
executed per second.
A lower TCR value implies a higher throughput.

\paragraph{Individual execution times}
To facilitate deeper analyis, the benchmark driver also collects all individual query execution times.
Based on these, the benchmark reports must include statics for each operation type (min, max, mean, $P_{50}$, $P_{90}$, $P_{95}$, and $P_{99}$ of the execution times).

\paragraph{Driver implementation in v2}
The \interactivevtwo is implemented in Java~17.
It consists of \numprint{26500} lines of code for the core project and an additional \numprint{18000} lines of test code.
The new version contains several patches including bug fixes, usability improvements, and performance optimizations.%
\footnote{\tt \href{https://github.com/ldbc/ldbc_snb_interactive_driver/releases/tag/v2.0.0-RC2}{github.com/ldbc/ldbc\_snb\_interactive\_driver/releases/tag/v2.0.0-RC2}}

\section{Parameter curation}
\label{sec:parameter-curation}

To prevent caching query results, the \snbinteractivevtwo driver instantiates the parameterized complex read (\CR) query templates with different \emph{substitution parameters} (\aka parameter bindings).
However, as explained in \autoref{sec:simplicity},
the na\"ive approach (using a uniform random sampling of parameters and ignoring updates)
leads to unstable runtimes,
which compromise both the benchmark's understandability and reproducibility.
To ensure stable runtimes, LDBC invented \emph{parameter curation} techniques, which select parameters that produce query runtimes with a unimodal (preferably Gaussian) distribution~\cite{DBLP:conf/tpctc/GubichevB14,DBLP:journals/pvldb/SzarnyasWSSBWZB22}.

\subsection{Building blocks for parameter curation}

\paragraph{Temporal bucketing}
\label{sec:temporal-bucketing}
To ensure that operations are always executable, \ie they avoid targeting nodes that are yet to be inserted or ones that are already deleted, the parameter curation process in \interactivevtwo employs \emph{temporal bucketing}.
Namely, we create a parameter bucket for \emph{each day in the simulation time of the update streams},
\ie each day in the simulation time has its own distinct set of parameters.
This is a novel feature in \interactivevtwo{} -- previous SNB benchmarks lacked this feature and only selected parameters from the \emph{initial snapshot}.

\paragraph{Factor tables}
As shown in \autoref{fig:interactive-components}, the parameter generation is a two-step process.
The \emph{factor generator} produces \emph{factor tables}, which contain data cube-like summary statistics~\cite{DBLP:journals/datamine/GrayCBLRVPP97} of the temporal graph such as the number of \Messages for friends.
The factor generator is executed in a distributed setup using Spark as this computation includes expensive joins over large tables,
\eg $\knows(\snbperson, \snbfriend) \bowtie \hasCreator(\snbperson, \snbcomment)$.

\subsection{Parameter curation for relational queries}

For relational queries (without path-finding), we based our parameter generation on two techniques.

\paragraph{(1) Selecting windows}
To select the parameters that are expected to yield similar runtimes, we look for windows with the smallest variance for a given value using SQL window functions.
The parameters are first sorted and grouped together based on their difference in frequency.
Groups that are smaller than a given minimum threshold are discarded to select a group of
parameters large enough to generate a sufficient amount of parameters.
From the latter, we select
the group with the smallest standard deviation.

\paragraph{(2) Selecting distributions}
For queries where we want to select parameters that are correlated or anti-correlated, we use factor tables encoding possible combinations (\eg \texttt{countryPairsNumFriends} for \CR[3]):
we select values near a high percentile for the correlated and a low percentile for the anti-correlated case.

\paragraph{Generating the parameters}
The parameter candidates discovered by the previous approaches are stored in temporary tables.
The parameter generation step uses these tables to select parameters for each day in the update stream.

\subsection{Parameter curation for path-finding queries}
\label{sec:path-curation}

\paragraph{The effect of deletes}
A key distinguishing feature of graph data management systems is their first-class support for path queries~\cite{DBLP:journals/csur/AnglesABHRV17}.
We demonstrate why ensuring stable query runtimes for path queries is particularly challenging through the example of \autoref{fig:paths}, where we query for the (unweighted) shortest path between \emph{Ada} and \emph{Bob} over a dynamic graph.
Initially, at $t = 1$, the length of the shortest path is 4~hops.
Then, the edge between \emph{Carl} and \emph{Dan} is deleted, making \emph{Ada} and \emph{Bob} unreachable from each other at $t = 2$.
Finally, a new edge is inserted between \emph{Carl} and \emph{Bob}, yielding a shortest path of length 3 at $t = 3$.
This illustrates how a given input parameter (a pair of \Persons) can oscillate between being reachable and being in disjoint connected components over a short period.
To ensure stable query runtimes for path queries in the presence of inserts and deletes, \interactivevtwo introduces a novel \emph{path curation} algorithm, which produces pairs of \Person nodes whose shortest path length from each other is guaranteed to be exactly $k$ hops at any point during a given day.

\paragraph{Graph construction}
The parameter curation algorithm builds two variants of the \Person--\knows--\Person subgraph for each day based on the \emph{temporal graph}:
graph $G_1$ has the inserts applied until the beginning of the day and the deletes applied until the end of the day,
while $G_2$ has the deletes applied until the beginning of the day and the inserts applied until the end of the day.
For a given pair of \Person nodes, their shortest path length in $G_1$ is an upper bound $k_\mathrm{upper}$ on their shortest path length at any point in the day -- when the inserts during the day are gradually applied, the shortest path length can only become shorter.
Conversely, $G_2$ gives a lower bound $k_\mathrm{lower}$ for the shortest path -- the deletes can only make the shortest path length become longer.

\paragraph{Parameter selection}
The bounds provided by $G_1$ and $G_2$ guarantee for the shortest path length $k$ that $k_\mathrm{lower} \leq k \leq k_\mathrm{upper}$ will hold at any point during the day.
We can ensure that $k$ will stay constant during the day by selecting \Person pairs where $k_\mathrm{lower} = k_\mathrm{upper}$ holds.
To this end, we select pairs who are exactly 4~hops apart in both $G_1$ and $G_2$, hence they will be always~4 hops apart during the given day.
Unreachable pairs of nodes can be generated by calculating the connected components of $G_2$ and selecting nodes from disjoint components.
The path curation for both the reachable and the unreachable cases is implemented using the NetworKit graph algorithm library~\cite{lit:networkit}.

\newsavebox{\largestimage}
\begin{figure}[htb]
    \vspace{-2.5ex}
    \savebox{\largestimage}{\includegraphics[width=.61\textwidth]{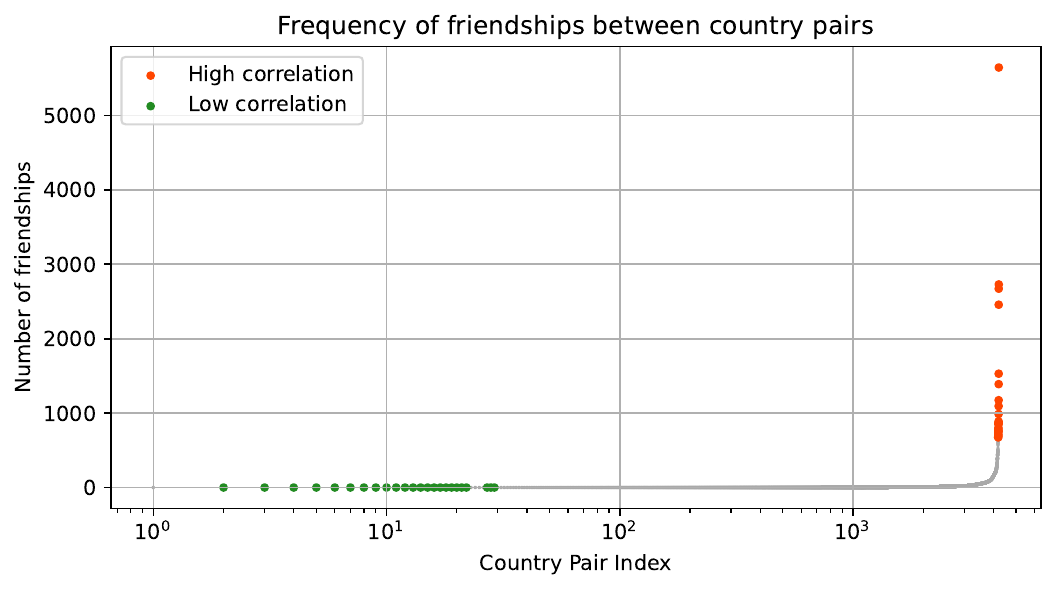}}%
    \begin{subfigure}{0.36\textwidth} 
        \centering
        \raisebox{\dimexpr.5\ht\largestimage-.5\height}{%
            \includegraphics[width=\textwidth]{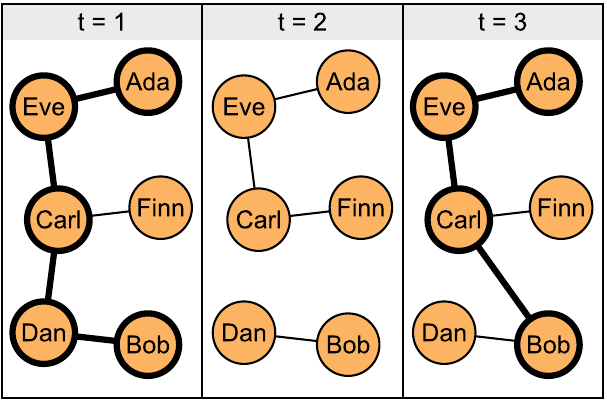}
        }
        \caption{
            Shortest path (denoted with thick lines) between \emph{Ada} and \emph{Bob} in the presence of updates.
        }
        \label{fig:paths}
    \end{subfigure}
    \hspace{2mm}
    \begin{subfigure}{0.61\textwidth}
        \centering
        \includegraphics[width=\linewidth]{figures/appendix/paramgen-frequency-disc-countries}
        \vspace{-3ex}
        \caption{
            Pairs of \Countries in the \texttt{countryPairsNum\-Friends} factor table representing the number of friendships between both \Countries.
        }
        \label{fig:paramgen-frequency-disc-countries}
    \end{subfigure}
    \caption{Example graph and distribution for path curation.}
    \label{y}
    \vspace{-4.5ex}
\end{figure}

\subsection{Query variants}
\label{sec:query-variants}

The new workload introduces variants for three queries:
\CR[3], \CR[13], and \CR[14].

\paragraph{\CRInTitle[3]: Correlated \vs anti-correlated Countries}
\label{sec:cr3-variants}
We introduce variants for \CR[3] (\autoref{fig:interactive-complex-read-03}):
variant \CR[3(a)] starts from \Countries that have a high correlation in the friendship network,
while
variant \CR[3(b)] starts from \Countries that have a low correlation of friendships between.
To generate these inputs, we use the \texttt{countryPairsNumFriends} factor table visualized in \autoref{fig:paramgen-frequency-disc-countries} and select values at percentile~1.00 for variant~\texttt{(a)} and percentile~0.01 for variant~\texttt{(b)}.

\paragraph{\CRInTitle[13] and \CRInTitle[14]: Reachable \vs unreachable Persons}
Path queries are expected to have different runtimes if there is a path \vs when there is no path.
While the performance characteristics vary highly between systems, in principle, the ``no path'' case should be simpler in the SNB graph, where one of the nodes is always in a small connected component.
To distinguish between these cases, we have two variants for the two path queries \CR[13] and \CR[14].
For variants~\snbOperation{(a)} we select \Person pairs which \emph{do not have a path},
and for variants~\snbOperation{(b)} we select pairs which \emph{have} a path of length~4.

\subsection{Parameter generator implementation}
\label{sec:paramgen-implementation}

The parameter generator is implemented in Python using NetworKit~\cite{lit:networkit} and SQL queries executed by DuckDB~\cite{DBLP:conf/sigmod/RaasveldtM19}.
Based on our experiments in~\cite[Figure~4.3]{david-puroja-msc}, the new parameter generator is scalable.
Even with the significant extra work performed for temporal bucketing,
it outperforms the old parameter generator by more than $100\times$ on SF\OneK,
and finishes in less than 1.5 hours on SF\TenK.

\section{Using the \snbinteractivevtwo workload}
\label{sec:framework}

In Sections~\ref{sec:data} to~\ref{sec:parameter-curation}, we presented the components that make up the SNB \interactivevtwo benchmark (\autoref{fig:interactive-components}):
its data sets, operations, driver, and parameter generator.
We continue by describing how the benchmark is used, including its current implementations and considerations for auditing implementations.

\subsection{Implementations}
\label{sec:implementations}

The portability of \interactivevtwo is demonstrated by having four complete initial implementations%
\footnote{\url{https://github.com/ldbc/ldbc_snb_interactive_impls}}
based on the Neo4j graph DBMS; and the Microsoft SQL Server, PostgreSQL, and Umbra RDBMSs.
The Neo4j implementation uses the Cypher query language~\cite{DBLP:conf/sigmod/FrancisGGLLMPRS18}.
SQL Server uses the Transact-SQL language with the graph extension.%
\footnote{\url{https://learn.microsoft.com/en-us/sql/relational-databases/graphs/sql-graph-overview?view=sql-server-ver16}}
PostgreSQL and Umbra both use SQL's PostgreSQL dialect.
All of these implementations passed cross-validation against each other.

\subsection{Auditing}
\label{sec:auditing}
\label{sec:fdr}

LDBC's benchmarks come with stringent auditing guidelines to ensure that they are implemented correctly and the results derived during benchmark runs are reproducible.
Audits are carried out by independent auditors who are certified by the benchmark task force.
The auditor ensures that an implementation is compliant with the LDBC specification by performing a thorough code review, running ACID tests, and executing the benchmark.
The results of audits are published as \emph{full disclosure reports} and systems are ranked on the LDBC website according to their throughput.%
\footnote{\url{https://ldbcouncil.org/benchmarks/snb-interactive/}}
In the following, we highlight important aspects of the \snb auditing guidelines such as rules for precomputation and ACID tests.
For the detailed auditing guidelines, we refer the reader to the \snb specification~\cite[Chapter 7]{DBLP:journals/corr/abs-2001-02299}.

\paragraph{Precomputation}
The auditing guidelines permit the use of precomputed auxiliary data structures (views, indexes, views, \etc) provided that they are always kept up-to-date upon update operations.
A frequent use of precomputations is the creation of a \rootPost edge for each \Message,
which points to the root \Post of the \Message's thread.
Implementers may decide to store information redundantly, \eg by adding a property to the \Forum--\hasMember--\Person edge that contains the number of \Posts in the \Forum, for improved locality during query execution.

\paragraph{ACID tests}
To ensure thorough testing of transactional guarantees,
\snbinteractivevtwo uses a separate ACID suite~\cite{DBLP:conf/tpctc/WaudbySKMBS20},
which tests for 10~transactional anomalies.
While \interactivevone only requires systems to guarantee the \emph{read committed} isolation level,
the inclusion of delete operations necessitates \emph{snapshot isolation} to ensure queries read a consistent database state.
To illustrate this consider a graph with four nodes $n_1, n_2, n_3, n_4$, and three edges $n_1 \rightarrow n_2 \rightarrow n_3 \rightarrow n_4$.
Assume transaction $T_a$ begins traversing from $n_1$, reading $n_1$, $n_2$, and $n_3$. 
Then, $T_b$ deletes $n_2$ and commits.
Then, $T_c$ inserts $n_5$, connecting $n_3$ and $n_4$, ($n_3 \rightarrow n_5 \rightarrow n_4$), and commits. 
$T_a$ then reads $n_5$ and (incorrectly) concludes that $n_4$ is reachable from $n_1$ -- when in fact at no point in time was this a valid database state.
The ACID tests also include a \emph{durability test:}
during a benchmark run, the system-under-test is shut down abruptly and restarted afterward.
The system is expected to guarantee durability, which is verified by the auditor who checks whether the last update operations issued by the driver are reflected in the database's state after recovery.

\paragraph{Full disclosure report (FDR)}
Audited benchmark results must be accompanied by a full disclosure report (FDR).
The FDR documents the benchmark setup for reproducibility and contains the detailed results of the benchmark run
(including statistics of the individual query runtimes).

\section{Related work: LDBC FinBench}
\label{sec:related-work}
\label{sec:finbench}

The LDBC FinBench (Financial Benchmark) targets distributed scale-out transactional graph database management systems.
It is set in the financial domain and uses concepts such as \textsf{Account}, \textsf{Loan}, and \textsf{transfer}.
Its data distribution follows the characteristics of the financial domain, where a hub vertex (\eg a large e-commerce vendor) may have billions of edges.
To make queries tractable on such vertices, the workload employs \emph{truncation}, \ie a traversal only uses a truncated set of edges, \eg the \numprint{5000} most recent edges.
A requirement directly derived from the financial domain is having \emph{strict latency requirements} for some queries, \eg 99\% of a given query's executions have to finish in 100~ms. %
The workload also includes path-finding queries that can be expressed as \emph{regular queries with memory}~\cite{DBLP:conf/icdt/LibkinV12}.

\section{Conclusion}
\label{sec:conclusion}

\paragraph{Summary}
In this paper, we summarized the LDBC SNB \interactivevone workload and explained its shortcoming to motivate its renewal.
We then presented the draft version of the \interactivevtwo workload, which is expected to be very close to the final version of the workload.
The new workload uses a data generator producing deep cascading delete operations, includes a completely reworked driver and workload scheduler, and a scalable parameter generator.
We compared the workload against other benchmarks and highlighted its key novel features that allow the incorporation of delete operations while keeping important guarantees such as stable query runtimes.
While the benchmark was substantially reworked,
we made an effort to keep the user-facing changes minimal
and only replaced a single read query, \CR[14].
We believe users with an existing v1 implementation can adopt the new version with reasonable development cost and extend their experiments to use larger scale factors in a matter of days.
Therefore, we expect users to quickly migrate to the new version upon its release.

\paragraph{Future work}
As the next step in the \interactivevtwo workload's development process, the SNB task force will finalize the workload, conduct \emph{standard-establishing audits} on two reference implementations and submit the workload for acceptance by the LDBC Members Policy Council.
Audits are expected to commence in 2024.
The task force will keep maintaining the workload in the coming years.
In future versions of \snbinteractive, we plan to incorporate long-running transactions, schema constraints~\cite{DBLP:conf/sigmod/AnglesBDFHHLLLM21}, and regular path queries~\cite{DBLP:conf/icdt/LibkinV12}.

\section*{Acknowledgements}

{\footnotesize
We would like to thank our collaborators who contributed with feedback and implementations to the SNB Interactive v2 workload:
Altan Birler,
Arvind Shyamsundar,
Benjamin A. Steer,
and
Dávid Szakállas.
Jack Waudby was supported by the Engineering and Physical Sciences Research Council, Centre for Doctoral Training in Cloud Computing for Big Data [grant number EP/L015358/1].
}

\bibliographystyle{abbrv}
\bibliography{ms}

\end{document}